\documentclass[11pt]{article}

\usepackage{amsmath}
\usepackage{amssymb}
\usepackage{epsfig}
\usepackage[bf,small]{caption}

\newcommand{\bea}{\begin{eqnarray*}}
\newcommand{\eea}{\end{eqnarray*}}

\newcommand{\Ap}{\hat{A}}
\newcommand{\CPM}{\hat{C}}
\newcommand{\gcc}{\mathrm{g}/\mathrm{cm}^3}

\newcommand{\GeV}{\,\mathrm{GeV}}

\newcommand{\eV}{\,\mathrm{eV}}

\newcommand{\dm}[1]{{\Delta m^2_{#1}}}

\DeclareMathOperator{\im}{Im}
\DeclareMathOperator{\re}{Re}

\begin{document}


\begin{titlepage}

\renewcommand{\thefootnote}{\alph{footnote}}

\ \vspace*{-3.cm}
\begin{flushright}
  {\hfill TUM--HEP--404/01}\\
  {\ }
\end{flushright}

\vspace*{0.5cm}

\renewcommand{\thefootnote}{\fnsymbol{footnote}}
\setcounter{footnote}{-1}

{\begin{center}
{\Large\bf Analytic Approximations for Three Neutrino Oscillation Parameters and Probabilities in Matter}

\end{center}}
\renewcommand{\thefootnote}{\alph{footnote}}

\vspace*{.8cm}
{\begin{center} {\large{\sc
                Martin~Freund
                }}
\end{center}}
\vspace*{0cm}
{\it 
\begin{center}  
Theoretische Physik, Physik Department, 
Technische Universit\"at M\"unchen,\\
James--Franck--Strasse, 85748 Garching, Germany

Email: martin.freund@physik.tu-muenchen.de  
\end{center} }

\vspace*{1.5cm}

{\Large \bf 
\begin{center} Abstract \end{center} }
The corrections to neutrino mixing parameters in the presence
of matter of constant density are calculated systematically as series 
expansions in terms of the mass hierarchy $\dm{21}/\dm{31}$. The parameter mapping 
obtained is then used to find simple, but nevertheless accurate formulas 
for oscillation probabibilities in matter including CP-effects. Expressions 
with one to one correspondence to the vacuum case are derived, 
which are valid for neutrino energies above the solar resonance energy. 
Two applications are given to show that these results are a useful and powerful 
tool for analytical studies of neutrino beams passing through the Earth mantle or core:
First, the ``disentanglement problem'' of matter and CP-effects in the CP-asymmetry
is discussed and second, estimations of the statistical sensitivity to the CP-terms 
of the oscillation probabilities in neutrino factory experiments are presented.

\vspace*{.5cm}

\end{titlepage}

\newpage

\renewcommand{\thefootnote}{\arabic{footnote}}
\setcounter{footnote}{0}


%

\section{Introduction \label{sec:SEC-intro}}

With the development of long baseline neutrino beams passing through the mantle 
of the Earth, three flavor neutrino oscillation with a constant matter profile is presently
drawing attention. Some effort has been spared on the exact solution of the
connected cubic eigenvalue problem \cite{EXACT}. However, the obtained
solutions are huge and were up to now only used in computer based calculations. 
Also approximative solutions for oscillation probabilities and mixing angles have been 
proposed for several parameter regions \cite{APPROX}, which are interesting
and useful. The intention of this work is to first derive analytic approximations
for the mixing parameters in matter\footnote{
Oscillation in matter can be described by a mapping
of the six basic parameters $\theta_{12}$, $\theta_{13}$,
$\theta_{23}$, $\dm{21}$, $\dm{31}$, and $\delta$ similar to 
the well-known two neutrino oscillation formulas in matter.}
according to the standard parameterization, which then allows to compute all 
desired quantities 
like probabilities or amplitudes from the known expressions in vacuum by 
substitution. 
The parameters in matter are calculated in a series expansion in the small mass 
hierarchy parameter $\alpha := \dm{21}/\dm{31}$. The obtained results are discussed 
and then applied to the appearance channel probability $P(\nu_e \rightarrow \nu_\mu)$. 
A simple solution, which is easy to use, but nevertheless 
accurate over a wide parameter range is obtained.
No new notation is introduced besides
abbreviations known from two neutrino oscillation in matter. Furthermore, the result 
shows at first sight the convergence to the vacuum case at small baselines and thus is 
directly connected to the terms in vacuum. 
The approximate solutions obtained with this method are a powerful tool for further 
analytical studies. To demonstrate this, two
applications are given. First the derived expressions are exploited to
compute the frequently used quantity called the CP-asymmetry  $A^\mathrm{CP}$, 
which has considerable importance in CP-violation studies. The problem is that matter 
effects cause contributions to the CP-asymmetry, which cannot easily be distinguished 
from intrinsic CP-effects. Here, expressions for $A^\mathrm{CP}$ in matter are
given for high neutrino energies (more precise: low $L/E_\nu$). The result is then
used to investigate what can be learned from the energy dependence of $A^\mathrm{CP}$.
The second application given estimates the statistical sensitivity to the CP-terms
of the oscillation probabilities in neutrino factory long baseline experiments. Plots
are presented, which show the magnitude of CP-effects at different baselines
and beam energies. Contrarily to what presently can be found in the literature, the 
here obtained results indicate strongly that, in general, the low energy option is
not the best solution to measure effects from the CP-phase $\delta$. The reason
for this discrepancy is discussed. 

Throughout this work two assumptions will be made: First, that the mass hierarchy 
parameter $\alpha := \dm{12}/\dm{31}$, which is used as expansion parameter,
is small. Consider for example an atmospheric $\dm{}$ of $3.2 \cdot  10^{-3} \eV^2$ \cite{ATM} .
For solar mass differences of LMA-scale\footnote{The abbreviation ``LMA'' stands for 
Large Mixing Angle MSW-solution to the solar neutrino problem. The MSW-solution 
assumes resonance enhanced oscillation of
neutrinos passing the core of the sun.} \cite{SOLAR} between $10^{-5} \eV^2$ and 
$10^{-4} \eV^2$, $\alpha$ varies between 0.0031 and 0.031. 
Second, it will
be assumed that the mixing angle $\theta_{13}$ is small as indicated by reactor, solar,
and atmospheric experiments. The strongest bound is given by the CHOOZ experiment
\cite{CHOOZ} with $\sin^2 2\theta_{13} < 0.1$. 
The smallness of this parameter
will be used to classify terms, which appear in the expressions for oscillation
probabilities. 
The mixing angles $\theta_{12}$ and $\theta_{23}$ should be chosen from the
interval $[0,\pi/2]$.

\section{Three neutrino oscillation in vacuum}

In vacuum, the neutrino oscillation probabilities are given by the well-known
formulas
\begin{equation}
\label{3NFORMULA}
P( \nu_{e_l} \rightarrow \nu_{e_m} )
= \delta_{lm} - 4\sum_{i>j} \mathrm{Re} J_{ij}^{lm}\sin^2\hat{\Delta}_{ij} - 
2\sum_{i>j} \mathrm{Im} J_{ij}^{lm}\sin 2\hat{\Delta}_{ij} ~,
\end{equation}
with the abbreviations $J_{ij}^{lm} := U_{li}U_{lj}^*U^*_{mi}U_{mj}$ and 
$\hat{\Delta}_{ij} := \Delta m^2_{ij}L/(4E)$. Here, $U$ is the mixing matrix 
of the neutrino sector in standard parameterization form:
\begin{equation}
U = \left(
\begin{array}{ccc} 
c_{12} c_{13} & 
c_{13}s_{12} & 
 e^{-i\delta}s_{13} 
\\
-s_{12}c_{23} - e^{i\delta} c_{12}s_{13}s_{23} &
c_{12}c_{23} -  e^{i\delta} s_{12}s_{13}s_{23} &
c_{13}s_{23}  
\\ 
- e^{i\delta} c_{12}s_{13}c_{23} + s_{12}s_{23} &
- e^{i\delta} s_{12}s_{13}c_{23} - c_{12}s_{23} &
c_{13}c_{23} 
\end{array}
\right) ~.
\end{equation}
Since in this work, the hierarchy $|\dm{21}| \ll |\dm{31}|$ between the two 
 mass squared differences is exploited, from now on all mass squared differences
 will always be related to the atmospheric squared mass difference: 
$\dm{31} =: \Delta$, $\dm{21} = \alpha \Delta$, $\dm{32} = (1-\alpha) \Delta$,
and $\hat\Delta =  \Delta L/(4E) $. Series expansion up to 
order $\alpha^2$ gives the following important terms in the oscillation
probability $P(\nu_e \rightarrow \nu_\mu) \approx P_0 + P_{\sin \delta} + P_{\cos \delta} + P_3$:
\begin{subequations}
\label{PROBVACUUM}
\begin{eqnarray}
P_0 &=& \sin^2 \theta_{23} \sin^2 2\theta_{13} \sin^2 \hat{\Delta}  \\
P_{\sin\delta} &=&  \alpha\; \sin\delta \cos\theta_{13} \sin 2\theta_{12} \sin 2\theta_{13} \sin 2\theta_{23}
\sin^3\hat{\Delta}  \\
P_{\cos\delta} &=&  \alpha\; \cos\delta \cos\theta_{13} \sin 2\theta_{12} \sin 2\theta_{13} \sin 2\theta_{23}
 \cos \hat{\Delta} \sin^2 \hat{\Delta}   \\
P_3 &=& \alpha^2 \cos^2 \theta_{23} \sin^2 2\theta_{12} \sin^2 \hat{\Delta} 
\end{eqnarray}
\end{subequations}
Expanding the oscillatory terms in $\alpha$ means linearization of the
oscillation over the solar mass squared difference. This gives valid 
results only for $\alpha \hat{\Delta} \lesssim 1$. With todays knowledge about
neutrino masses this does not cause crucial errors for neutrino energies
above 1~GeV at baselines below approximately 10000~km. The two terms  
$P_{\sin\delta}$ and $P_{\cos\delta}$,
containing the CP-phase $\delta$, are both of
order $\alpha$ and hence suppressed 
by the mass hierarchy. This reflects the fact that CP-effects vanish when
the mass hierarchy becomes large. Besides the factor $\sin^2 \theta_{23}$, the 
term $P_0$ is similar to the two neutrino oscillation probability which in matter
is expected to show the resonant behavior called MSW-effect \cite{MSW}. The term
$P_3$ is the
only term of order $\alpha^2$, which is not suppressed by the small mixing angle
$\theta_{13}$. Hence, it is important to take  this term into account
when $\theta_{13}$ is small. If $\theta_{13}$ is not too far away from the 
CHOOZ-bound, $P_3$ can safely be neglected.
All other terms of order $\alpha^2$ are additionally suppressed by one or more
powers of $\theta_{13}$ and are not listed here.

\section{Mixing parameters in matter}

In matter, the effective Hamiltonian in flavor basis is given by
\begin{equation}
\mathcal{H} =  \frac{1}{2E} \left[ U \left( 
\begin{array}{ccc} 
m_1^2 & 0 & 0 \\ 
0 & m_2^2 & 0 \\ 
0 & 0 & m_3^2 
\end{array}
\right) U^\dagger + \left( \begin{array}{ccc} A & 0 & 0 \\ 0 & 0 & 0 \\ 0 & 0 & 0 \end{array} \right)
\right] ~.
\end{equation}
Here $U = U_{23}(\theta_{23}) U_{13}(\theta_{13},\delta)  U_{12}(\theta_{12})$ is
the mixing matrix, which rotates from mass to flavor basis. The second term is 
generated by matter effects with $A = 2 V E_\nu$ and $V = \sqrt{2} G_F n_e$, where
$G_F$ is the Fermi coupling constant and $n_e$ is the electron density of the
matter, which is crossed by the neutrino beam.

The matter term is invariant under rotations in the 23-subspace.
Separating $\mathrm{diag}(m_1^2, m_1^2, m_1^2)$ which, as global phase, does 
not contribute to the probability, and 
using the above defined parameters, the Hamiltonian can be written in the form 
\begin{equation}
\label{HAMILTONIANA}
\mathcal{H} =  \frac{\Delta}{2E} U_{23} \left[ U_{13} U_{12} \left( 
\begin{array}{ccc} 
0 & 0 & 0 \\ 
0 & \alpha  & 0 \\ 
0 & 0 & 1 
\end{array}
\right) U_{12}^\dagger U_{13}^\dagger  + \left( \begin{array}{ccc} \frac{A}{\Delta} & 0 & 0 \\ 0 & 0 & 0 \\ 0 & 0 & 0 \end{array} \right)
\right] U_{23}^\dagger ~.
\end{equation}
With
\begin{equation}
U_\delta := 
\left(
\begin{array}{ccc} 
1 & 0 & 0 \\ 
0 & 1 & 0 \\ 
0 & 0 & e^{i\delta} 
\end{array}
\right) ~,
\end{equation}
the relations 
\begin{subequations}
\begin{eqnarray}
U_\delta^\dagger \, U_{13}(\theta_{13},\delta) \, U_\delta &=& U_{13}(\theta_{13},0) ~,\\
U_\delta^\dagger \, U_{12}(\theta_{12}) \, U_\delta &=& U_{12}(\theta_{12}) ~,\\
U_\delta^\dagger \, \mathrm{diag}(a,b,c) \, U_\delta &=&  \mathrm{diag}(a,b,c)  
\end{eqnarray}
\end{subequations}
are valid.
Inserting the identity matrix $U_\delta \, U_\delta^\dagger$ at the appropriate 
places in eq.~(\ref{HAMILTONIANA}) gives
\begin{equation}
\label{HAMILTONIAN}
\mathcal{H} =  \frac{\Delta}{2E} U_{23}\, U_\delta \underbrace{\left[ U_{13}(\theta_{13},0)\, U_{12} \left( 
\begin{array}{ccc} 
0 & 0 & 0 \\ 
0 & \alpha  & 0 \\ 
0 & 0 & 1 
\end{array}
\right) U_{12}^\dagger \, U_{13}(\theta_{13},0)^\dagger  + \left( \begin{array}{ccc} \frac{A}{\Delta} & 0 & 0 \\ 0 & 0 & 0 \\ 0 & 0 & 0 \end{array} \right)
\right]}_{M} U_\delta^\dagger \, U_{23}^\dagger ~.
\end{equation}

Diagonalization of the real matrix $M$ by 
$\hat U :=  U_{23}(\hat\theta_{23}) \, U_{13}(\hat\theta_{13}) \, U_{12}(\hat\theta_{12})$
together with the part which was factored out gives the complete mixing matrix 
$U'$ in matter:
\begin{equation}
U' = U_{23}(\theta_{23}) \, U_\delta \, U_{23}(\hat\theta_{23}) \, U_{13}(\hat\theta_{13}) \, U_{12}(\hat\theta_{12}) ~.
\end{equation}

\subsubsection*{Mixing angles in standard parameterization form}

The matrix $U'$ must still be brought to the standard form. 
The matrix
\begin{equation}
U_{23}(\theta_{23}) \, U_\delta \, U_{23}(\hat\theta_{23}) = 
\left(
\begin{array}{ccc} 
1& 0 & 0 \\ 
0 & C & S \\ 
0 & -e^{i \delta}S^* &  e^{i \delta} C^* 
\end{array}
\right)
\end{equation}
with
\begin{subequations}
\begin{eqnarray}
C &:=& \cos \theta_{23} \cos \hat\theta_{23} - e^{i \delta} \sin \theta_{23} \sin \hat\theta_{23} ~, \\
S &:=& \cos \theta_{23} \sin \hat\theta_{23} + e^{i \delta} \sin \theta_{23} \cos \hat\theta_{23} 
\end{eqnarray}
\end{subequations}
can be made 
real by the phase rotations $\beta := -\mathrm{arg}\, C$, 
$\gamma := \mathrm{arg}\, S$, and
$\delta' := \mathrm{arg}\, C -\mathrm{arg}\, S$
\footnote{Using $|C|$ and $|S|$ in eq.~(\ref{trans}) further 
restricts the parameter space for $\theta_{23}$.
Since $\theta_{23}$ is assumed to be close to $\pi/4$ 
and $\hat\theta_{23}$ in general is small, this problem is not  
relevant for the calculations presented here.}
:
\begin{equation}
\label{trans}
\left(
\begin{array}{ccc} 
1& 0 & 0 \\ 
0 & e^{-i\beta} & 0 \\ 
0 & 0 &  -e^{(i\delta -\gamma)} 
\end{array}
\right) \,
U_{23}(\theta_{23}) \, U_\delta \, U_{23}(\hat\theta_{23}) \,
\left(
\begin{array}{ccc} 
1& 0 & 0 \\ 
0 & 1 & 0 \\ 
0 & 0 & e^{-i\delta'}  
\end{array}
\right) 
= 
\left(
\begin{array}{ccc} 
1& 0 & 0 \\ 
0 & |C| & |S| \\ 
0 & -|S| & |C|  
\end{array}
\right) ~.
\end{equation}
This gives
\begin{equation}
U' = \left(
\begin{array}{ccc} 
1& 0 & 0 \\ 
0 & e^{i\beta} & 0 \\ 
0 & 0 &  -e^{(i\gamma -\delta)} 
\end{array}
\right) \,
\left(
\begin{array}{ccc} 
1& 0 & 0 \\ 
0 & |C| & |S| \\ 
0 & -|S| & |C|  
\end{array}
\right) \, 
U_{\delta'} \, U_{13}(\hat\theta_{13}) \, U^\dagger_{\delta'} \, U_{12}(\hat\theta_{12}) \, U_{\delta'} ~.
\end{equation}
The phase rotations on the left 
and on the right can be absorbed in the field vectors, yielding then
$U'$ in standard parameterization form:
\begin{equation}
U' = U(\theta_{23}') \, U_{13}(\hat\theta_{13},\delta') \, U_{12}(\hat\theta_{12}) ~.
\end{equation}
This finally means, that the (standard) mixing angles 
$\theta_{13}'$ and $\theta_{12}'$ in matter are equal
to $\hat\theta_{13}$ and $\hat\theta_{12}$ which are 
obtained from the matrix that diagonalizes $M$. The
matter correction $\hat\theta_{23}$, however,
mixes with the CP-phase $\delta$:
\begin{subequations}
\label{rel}
\begin{eqnarray}
\theta_{13}' &=& \hat\theta_{13} ~, \\
\theta_{12}' &=& \hat\theta_{12} ~, \\
\sin^2 \theta_{23}' &=& \cos^2 \theta_{23} \sin^2 \hat\theta_{23} 
+ \sin^2 \theta_{23} \cos^2 \hat\theta_{23}
+ 2 \cos \delta  \sin\theta_{23} \cos\theta_{23}  \sin\hat\theta_{23}\cos\hat\theta_{23} ~, \\
\sin \delta' &=& \sin\delta \, \frac{\sin 2\theta_{23}}{\sin 2\theta_{23}'} ~.
\end{eqnarray}
\end{subequations}
Equation~(\ref{rel}d) was first found by S.~Toshev \cite{TOSHEV}. There, a different 
parameterization is used, which -- for oscillations -- is equivalent to the
standard parameterization. It is important to note that the results given up to 
here are exact results for three neutrino oscillation in matter and do not
presume that the mass hierarchy parameter is small.

\subsubsection*{Calculation of the eigenvalues and eigenvectors}

Hereafter $\Ap$ will be used as abbreviation for $\frac{A}{\Delta}$.
Diagonalization of the matrix $M$ leads to the oscillation parameters in matter. 
Note that $M$ does not include the parameters $\theta_{23}$ and $\delta$, which
have been factored out. This will simplify the calculation of the eigenvalues and 
eigenvectors of $M$ considerably:

\begin{equation}
M = \left( \begin{array}{ccc} 
 s^2_{13} + \Ap + \alpha c^2_{13}  s^2_{12} &  \alpha  s_{12} c_{12} c_{13} &  s_{13} c_{13} -\alpha  s_{13}c_{13} s^2_{12}\\
\alpha  s_{12} c_{12}c_{13} & \alpha c^2_{12} & -\alpha  s_{12}c_{12} s_{13} \\ 
 s_{13} c_{13} - \alpha  s_{13}c_{13} s^2_{12} & -\alpha  s_{12}c_{12} s_{13} & c^2_{13} + \alpha  s^2_{12} s^2_{13}
\end{array} \right) ~.
\end{equation}
The invariants of the cubic eigenvalue problem are given by
\begin{subequations}
\label{invariants}
\begin{eqnarray}
\label{EQ:INVARIANTS}
I_1 &=&  \mathrm{Tr}(M) = \lambda_1 + \lambda_2 + \lambda_3 = \nonumber \\ 
    &=&  \Ap + 1 + \alpha ~,  \\
I_2 &=&  \frac{1}{2}\left[\mathrm{Tr}(M)-\mathrm{Tr}(M^2)\right] = \lambda_1\lambda_2 +  \lambda_1\lambda_3 +  \lambda_2\lambda_3 = \nonumber  \\
    &=&  \Ap \cos^2\theta_{13} + \alpha + \alpha \Ap \left(\sin^2 \theta_{13} \sin^2 \theta_{12} + \cos^2 \theta_{12} \right) ~,   \\
I_3 &=&  \mathrm{Det}(M) =  \lambda_1\lambda_2\lambda_3 = \nonumber \\
    &=&  \alpha \Ap \cos^2\theta_{13} \cos^2\theta_{12} ~.  \
\end{eqnarray}
\end{subequations}
Solving this system of equations in a series expansion of $\alpha$ gives the 
eigenvalues
\begin{subequations}
\label{eigenvalues}
\begin{eqnarray}
\lambda_1 &=&  \frac{1}{2}(\Ap+1-\CPM) + \alpha \; \frac{(\CPM+1-\Ap\cos2\theta_{13})\sin^2\theta_{12}}{2 \CPM} + \mathcal{O}(\alpha^2) ~,  \\
\lambda_2 &=&  \alpha \cos^2\theta_{12} + \mathcal{O}(\alpha^2) ~,  \\
\lambda_3 &=&  \frac{1}{2}(\Ap+1+\CPM) + \alpha \; \frac{(\CPM-1+\Ap\cos2\theta_{13})\sin^2\theta_{12}}{2 \CPM} + \mathcal{O}(\alpha^2) ~, 
\end{eqnarray}
\end{subequations}
with 
\begin{equation}
\CPM = \sqrt{(\Ap-\cos 2\theta_{13})^2+\sin^2 2\theta_{13}} ~.
\end{equation}
Here, $\CPM$ is the same square root, which appears in the two neutrino 
matter formulas.

Calculating the eigenvectors of $M$ in order $\mathcal{O}(\alpha)$ 
gives:
\begin{subequations}
\begin{eqnarray}
v_1 &=& \left( 
\begin{array}{c} 
\frac{\sin2\theta_{13}}{\sqrt{2\CPM(\Ap+\CPM-\cos2\theta_{13})}}
- \frac{\alpha\;\Ap \sin^2\theta_{12}\sin^2 2\theta_{13}}
{2 \CPM \sqrt{2 \CPM^2 (-\Ap + \CPM  + \cos 2\theta_{13})}} 
\\ 
\frac{\alpha\;(1 +\Ap - \CPM)\sin2\theta_{12}\sin\theta_{13}}
{(1 + \Ap + \CPM)\sqrt{2\CPM (\Ap + \CPM - \cos2\theta_{13})}} 
\\
-\frac{\sin2\theta_{13}}{\sqrt{2\CPM(-\Ap+\CPM + \cos2\theta_{13})}}
- \frac{\alpha\;\Ap \sin^2\theta_{12}\sin^2 2\theta_{13}}
{2 \CPM \sqrt{2 \CPM^2 (\Ap + \CPM  - \cos 2\theta_{13})}} 
\end{array}\right) + \mathcal{O}(\alpha^2)  ~, \\
v_2 &=& \left(\begin{array}{c} 
-\frac{\alpha\;\cos\theta_{12}\sin\theta_{12}}{\Ap \cos\theta_{13}}
\\ 
1
\\
\frac{\alpha\;(1 + \Ap) \cos\theta_{12} \sin\theta_{12} \sin\theta_{13}}{\Ap \cos^2\theta_{13}}
\end{array}\right) + \mathcal{O}(\alpha^2) ~,  \\
v_3 &=& \left(
\begin{array}{c}
\frac{\sin2\theta_{13}}{\sqrt{2\CPM(-\Ap+\CPM + \cos2\theta_{13})}}
+ \frac{\alpha\;\Ap \sin^2\theta_{12}\sin^2 2\theta_{13}}
{2 \CPM \sqrt{2 \CPM^2 (\Ap + \CPM  - \cos 2\theta_{13})}}
\\ 
\frac{\alpha\;(1 +\Ap - \CPM)\sin2\theta_{12}\sin\theta_{13}}
{(1 + \Ap + \CPM)\sqrt{2\CPM (-\Ap + \CPM + \cos2\theta_{13})}} 
\\ 
\frac{\sin2\theta_{13}}{\sqrt{2\CPM(\Ap+\CPM-\cos2\theta_{13})}}
- \frac{\alpha\;\Ap \sin^2\theta_{12}\sin^2 2\theta_{13}}
{2 \CPM \sqrt{2 \CPM^2 (-\Ap + \CPM  + \cos 2\theta_{13})}} 
\end{array}\right) + \mathcal{O}(\alpha^2)  ~.
\end{eqnarray}
\end{subequations}
There is one major problem concerning the calculation of the
eigenvalues and eigenvectors, which has to be addressed. Throughout the above series
expansion $\Ap$ was assumed to be different from zero. This is important 
as the results given above do not hold for $\Ap = 0$ in which case a
different series expansion in $\alpha$ would be obtained. This is a general
and important fact. In principle, it is also possible to give results
for small values of $|\Ap|$, which, however, would fail for larger $|\Ap|$.
The reason for this is that there are two different resonances occurring.
One for $\Ap = \alpha$ (solar resonance) and one for $\Ap = \cos 2\theta_{13}$ 
(atmospheric 
resonance). Each resonance produces a level-crossing of the eigenvalues.
To describe both level-crossings, the correct expression for the
eigenvalues are necessary. Being interested in approximative solutions,
one has to distinguish the two above mentioned cases. In this work the focus is
on the case $|\Ap| > \alpha$, which is appropriate for neutrino beams
above 1~GeV in matter densities of $2.8~\gcc$ (Earth mantle) or more. 
However, one must not expect that the expressions for the mixing parameters 
in matter will show the correct convergence for $\Ap \rightarrow 0$. For
$\dm{21} = 10^{-4} \eV^2$ and $2.8~\gcc$ we find that $\Ap > \alpha$ is
valid for $E_{\nu} > 0.5 \GeV$. This lower bound on the neutrino energy
decreases linearly with $\dm{21}$.

That the results for the eigenvalues and eigenvectors obtained from the series 
expansion are not good at the resonance $\Ap \approx 1$ is another point to mention. 
However, this does not have a crucial implication on the obtained results for 
the parameter mapping and oscillation probabilities. This issue will be
discussed later, at the appropriate places.  
 
\subsection*{Construction of \boldmath$\hat U$}

It is now possible to construct $\hat U$ from the 
eigenvectors $v_1$, $v_2$, and $v_3$. For this it is necessary to correctly
identify the order and the signs of the eigenvectors. In order to avoid 
divergences in the expressions for the mixing angles, it is appropriate to
change the order at the resonance $\Ap = \cos 2\theta_{13}$\footnote{
Another strategy would be to chose the order in such a way that in the
limit $|\Ap| \rightarrow 0$, the correct mixing matrix in vacuum is obtained.
However, since the expressions for the eigenvectors and eigenvalues are not good in
this limit, this is not a feasible solution here.}
:
\begin{equation}
\label{EQ:ORDERING}
\hat U = \left\{ \begin{array}{crl} 
\left( v_1 \, v_2 \, v_3 \right)^T & \mathrm{for} & \qquad \Ap < \cos 2\theta_{13}   \\
\left( v_3 \, v_2 \, v_1 \right)^T &  \mathrm{for} & \qquad \Ap > \cos 2\theta_{13}
\end{array}\right. ~.
\end{equation}
The second point is to bring $U'$ to a form which is consistent with the 
standard parameterization. This is not trivial and has to be carried 
out carefully for each of the different cases. As an example, the case
$\Ap < 0$ will be considered in detail:

As the vacuum angle $\theta_{23}$ was factored out from the beginning 
(eq.~(\ref{HAMILTONIAN})),
the matter induced change of this mixing angle $\hat\theta_{23}$ 
will be of order $\alpha$. This can be also seen by looking at the ($\mu,3$)-element 
of $\hat U$. Furthermore, by looking at the $(e,2)$-element, one finds that also 
$\hat\theta_{12}$ must be of order $\alpha$. Considering this with the replacements
$\hat s_{12} = \alpha \hat s_{12}^{(\alpha)}$,  
$\hat s_{23} = \alpha \hat s_{23}^{(\alpha)}$, and
$\hat s_{13} = \hat s_{13}^{(0)} + \alpha \hat s_{13}^{(\alpha)}$, one obtains the
following structure for $\hat U$:
\begin{equation}
\hat U = \left(
\begin{array}{ccc} 
\hat c_{13} & 
\alpha \hat c_{13}^{(0)} \hat s_{12}^{(\alpha)} & 
 \hat s_{13} 
\\
-\alpha (\hat s_{12}^{(\alpha)} +  \hat s_{13}^{(0)} \hat s_{23}^{(\alpha)}) &
1 &
\alpha \hat c_{13}^{(0)} \hat s_{23}^{(\alpha)}  
\\ 
- \hat s_{13} &
-\alpha( \hat s_{12}^{(\alpha)} \hat s_{13}^{(0)} + \hat s_{23}^{(\alpha)}) &
\hat c_{13}
\end{array}
\right) +\mathcal{O}(\alpha^2) ~.
\end{equation}
Then, $\sin \hat\theta_{13}$ and $\sin \hat\theta_{23}$ can be read off directly from 
$\hat U_{e3}$, $\hat U_{\mu3}$ and $\hat U_{\tau3}$: 
\begin{eqnarray}
\sin \hat\theta_{13} &=& \frac{\sin2\theta_{13}}{\sqrt{2\CPM(-\Ap+\CPM + \cos2\theta_{13})}}
+ \frac{\alpha\;\Ap \sin^2\theta_{12}\sin^2 2\theta_{13}}
{2 \CPM \sqrt{2 \CPM^2 (\Ap + \CPM  - \cos 2\theta_{13})}} +\mathcal{O}(\alpha^2) ~, \\
\sin \hat\theta_{23} &=&  \alpha \frac{(1 + \Ap - \CPM) \sin 2\theta_{12} \sin\theta_{13}}
{2(1 - \Ap + \CPM) \cos^2 \theta_{13}} +\mathcal{O}(\alpha^2)  ~.
\end{eqnarray}
To find $\sin \hat\theta_{12}$, it is now useful to split off $\hat\theta_{23}$.
The rest $\hat U = U_{23}^T(\hat\theta_{23})\,\hat U'$ should now be brought to 
the form
\begin{equation}
\left(
\begin{array}{ccc} 
\hat c_{13} & 
\alpha \hat c_{13}^{(0)} \hat s_{12}^{(\alpha)} & 
\hat s_{13} 
\\
-\alpha \hat s_{12}^{(\alpha)} &
1 &
0 
\\ 
-\hat s_{13}' &
-\alpha \hat s_{12}^{(\alpha)} \hat s_{13}^{(0)} &
\hat c_{13}
\end{array}
\right) +\mathcal{O}(\alpha^2) ~.
\end{equation}
The mixing angle  $\hat \theta_{12}$ can then be read off
from $\hat U'_{\mu1}$:
\begin{equation}
\sin \hat\theta_{12} = - \frac{\alpha\;\CPM\sin2\theta_{12}}
{\Ap \cos\theta_{13} \sqrt{2\CPM (-\Ap + \CPM + \cos2\theta_{13})}} +\mathcal{O}(\alpha^2) ~.
\end{equation}
\subsection*{Parameter mapping}
Considering the correct ordering of the eigenvectors (eq.~(\ref{EQ:ORDERING})) 
and following 
the above described steps, one can determine the complete parameter mapping for 
all regions of the $\Ap$ parameter space. 
Comprising, one obtains the following
expressions for the mixing parameters in matter:
\begin{subequations}
\label{MAPPING}
\begin{eqnarray}
\sin \theta'_{13} &=&  \frac{\sin2\theta_{13}}{\sqrt{2\CPM(\mp\Ap +\CPM \pm \cos2\theta_{13})}}
\pm \frac{\alpha\;\Ap \sin^2\theta_{12}\sin^2 2\theta_{13}}
{2 \CPM^2 \sqrt{2 \CPM (\pm\Ap + \CPM  \mp \cos 2\theta_{13})}} \\
\sin \theta'_{12} &=& 
\alpha \; \frac{\CPM \sin 2\theta_{12}}{|\Ap| \cos\theta_{13} \sqrt{2  \CPM(\mp\Ap + \CPM \pm \cos 2\theta_{13})}} \\ 
\sin \theta'_{23} &=&  \sin \theta_{23} + \alpha\, \cos\delta \;
\frac{\Ap \sin 2\theta_{12} \sin \theta_{13}\cos{\theta_{23}}}
{\pm 1 + \CPM \mp \Ap \cos 2\theta_{13}} \\
\sin \delta' &=&  \sin \delta \left( 1 - \alpha \, \frac{\cos\delta}{\tan 2\theta_{23}} \;
\frac{2 \Ap \sin 2\theta_{12} \sin \theta_{13} }
{\pm 1 + \CPM \mp \Ap \cos 2\theta_{13}}  \right) 
\end{eqnarray}
\end{subequations}
Here, in the expressions with choices for the sign, the upper sign holds for 
$\Ap < \cos 2\theta_{13}$ and the lower sign holds for $\Ap > \cos 2\theta_{13}$.
Higher orders than $\mathcal{O}(\alpha)$ are omitted.
To take into account also $\theta_{23}$ and $\delta$, which were factored out at 
the beginning, the equations~(\ref{rel}a-d) were applied. The expansion of
$\sin \delta'$ given here does not hold for $\theta_{23} \rightarrow 0$. 

From this parameter mapping it is possible to derive the
following quantities:
\begin{subequations}
\label{MAPPING2}
\begin{eqnarray}
\sin^2 2\theta'_{13} &=&   \frac{\sin^2 2\theta_{13}}{\CPM^2}  
+ \alpha \; \frac{2\Ap (-\Ap + \cos 2\theta_{13}) \sin^2\theta_{12}\sin^2 2\theta_{13}}{\CPM^4} \\
\sin 2\theta'_{12} &=& 
\alpha \; \frac{2\CPM \sin 2\theta_{12}}{|\Ap| \cos\theta_{13} \sqrt{2  \CPM(\mp\Ap + \CPM \pm \cos 2\theta_{13})}} \\ 
\sin 2\theta'_{23} &=&  \sin 2\theta_{23} + \alpha \cos\delta  \; 
\frac{2 \Ap \sin 2\theta_{12} \sin \theta_{13} \cos 2\theta_{23}}
{\pm 1 + \CPM \mp \Ap \cos 2\theta_{13}} 
\end{eqnarray}

\end{subequations}
For the mass squared differences one obtains:
\label{MASSDIFFERENCES}
\begin{equation}
({\Delta m^2_{21}}',{\Delta m^2_{31}}', {\Delta m^2_{32}}')  = \left\{ 
\begin{array}{ll} 
(\;\;\;\Delta m^2_3, \;\;\;\Delta m^2_2, \;\;\;\Delta m^2_1) & \mathrm{for} \quad \Ap < \cos 2\theta_{13} \\  
(-\Delta m^2_1,  -\Delta m^2_2, -\Delta m^2_3)  & \mathrm{for} \quad \Ap > \cos 2\theta_{13} 
\end{array} \right.
\end{equation}
with
\begin{subequations}
\label{MASSDIF}
\begin{eqnarray}
{\Delta m^2_1}' &:=&  \Delta (\lambda_3 - \lambda_2) \nonumber \\&=&  \frac{1}{2}(1+\Ap+\CPM) \Delta - \alpha \Delta \left( \cos^2\theta_{12} - \frac{(-1 + \CPM + \Ap\cos 2\theta_{13}) \sin^2\theta_{12}}{2\CPM} \right) ~, \\
{\Delta m^2_3}' &:=&  \Delta (\lambda_2 - \lambda_1) \nonumber \\&=&  \frac{1}{2}(-1-\Ap+\CPM) \Delta + \alpha \Delta \left( \cos^2\theta_{12} - \frac{(1 + \CPM - \Ap\cos 2\theta_{13}) \sin^2\theta_{12}}{2\CPM} \right) ~, \\
{\Delta m^2_2}' &:=&  \Delta (\lambda_3 - \lambda_1) \nonumber \\&=&  \CPM \Delta + \alpha \frac{\Delta (-1 + \Ap \cos 2\theta_{13}) \sin^2\theta_{12}}{\CPM}  ~. 
\end{eqnarray}
\end{subequations}

Looking at the
expressions for the mixing angles in matter, one obtains the following
interesting statements:

\subsection*{\boldmath $\mathbf{\sin^2 2\theta_{13}'}$}

In leading order, one finds the well-known resonant behavior of $\theta_{13}'$
familiar from two neutrino oscillation as MSW-resonance.
The order $\alpha$ correction to this leading result is suppressed
by two powers of $\theta_{13}$, and hence, is negligible small. A careful study
of the correction indeed shows that it is small and only important if 
precise results are to be obtained. The expressions for $\theta_{13}$ do not
show divergences for $|\Ap| \rightarrow 0$ and the vacuum limit is correctly 
described. Comparison with numerical results shows an excellent agreement even 
for $|\Ap| < \alpha$.

\subsection*{\boldmath $\mathbf{\sin 2\theta_{23}'}$}

In leading order, the mixing angle $\theta_{23}'$ is equal to the vacuum mixing 
angle $\sin 2\theta_{23}$.
The order $\alpha$ correction is double suppressed by $\theta_{13}$ and 
by $\cos 2\theta_{23}$ (when $\theta_{23}$ is close to $\pi/4$). 
Its proportionality to $\cos \delta$ is caused by the mixing of the
CP-phase $\delta$ with the $\mathcal{O}(\alpha)$ correction of $\theta_{23}'$
(eq.~(\ref{rel}c)). The expression for $\theta_{23}'$ shows the correct 
behavior for $|\Ap| \rightarrow 0$ and numerical results are consistent 
also for $|\Ap| < \alpha$.

\subsection*{\boldmath $\mathbf{\sin 2\theta_{12}'}$}

The quantity $\sin 2\theta_{12}'$ is of order $\alpha$. For $\alpha \rightarrow 0$ it does
not reproduce the vacuum parameter $\theta_{12}$. But this is not difficult to
understand. For $\alpha = 0$, the first term in the Hamiltonian (eq.~(\ref{HAMILTONIAN}))
is invariant under rotations in  the 12-subspace. This reflects the fact that
for $\alpha=0$ the solar mixing angle does not influence the oscillation
probabilities and could in principle be chosen arbitrarily. Interesting here
is that $\sin 2\theta_{12}'$, even for large values of $|\Ap|$, is proportional
to $\alpha$. In leading order of $\theta_{13}$ one finds that
$\sin 2\theta_{12}' = \alpha \; \sin 2\theta_{12} / |\Ap|$.
There appears a divergence for $|\Ap| \rightarrow 0$. The result is 
unphysical for $|\Ap| \lesssim \alpha$, which reflects the problem that the 
level crossing at the solar resonance is not correctly described. Since 
$|\Ap|$ is proportional to the neutrino energy $E_\nu$, $\sin 2\theta_{12}'$
is suppressed not only by the mass hierarchy, but also by large neutrino
energies.

\subsection*{\boldmath CP-phase $\mathbf{\delta}$}

The correction to the CP-phase $\delta$ in matter is triple suppressed by
the mass hierarchy $\alpha$, $\theta_{13}$, and $\tan^{-1} 2\theta_{23}$. 
For $\sin^2 2\theta_{23} = 1$, the CP-phase $\delta$ is not changed (in order $\alpha$).
The invariance of $\sin\delta \sin 2\theta_{23}$ under variations 
of the matter density $\rho$ (eq. (\ref{rel}d)) is an exact result, which
is independent from the approximations made.

\section{\boldmath CP-violation: $J_\mathrm{CP}$ in matter}

From the vacuum case it is known that the quantity $J_\mathrm{CP} = \im J_{ij}^{lm}$ 
drives the strength of CP-violating effects. In vacuum, it is given by
\begin{equation}
8 J_\mathrm{CP} = \sin \delta \cos \theta_{13} \sin 2\theta_{12} \sin 2\theta_{13} \sin 2\theta_{23} ~.
\end{equation}
Application of the parameter mapping (eqs.~(\ref{MAPPING})) gives $J_\mathrm{CP}'$ in matter:
\begin{eqnarray}
&& \sin \delta' \cos \theta_{13}' \sin 2\theta_{12}' \sin 2\theta_{13}' \sin 2\theta_{23}' = \nonumber \\
&& \frac{\alpha}{|\Ap| \CPM \cos^2 \theta_{13}} \sin \delta \cos \theta_{13} \sin 2\theta_{12} \sin 2\theta_{13} \sin 2\theta_{23}
 + \mathcal{O}(\alpha^2) ~.
\end{eqnarray}
One thus finds the important and simple result
\begin{equation}
J_\mathrm{CP}' = \frac{\alpha}{|\Ap| \CPM \cos^2 \theta_{13}} \, J_\mathrm{CP} ~.
\end{equation}
Applying this result to 
\begin{equation} 
J_\mathrm{CP}' \Delta m_{12}'  \Delta m_{31}' \Delta m_{32}' = J_{CP} \Delta^3 \alpha + \mathcal{O}(\alpha^2) ~,
\end{equation}
the Harrison-Scott invariance $J_\mathrm{CP}' \Delta m_{12}'  \Delta m_{31}' \Delta m_{32}' =  J_\mathrm{CP} \Delta m_{12}  \Delta m_{31} \Delta m_{32}$ 
\cite{HARRISONSCOTT} can be verified.

It is important to notice that also in matter all CP-violating effects are 
proportional to the mass hierarchy $\alpha$. In vacuum, the suppression of 
CP-effects through the mass hierarchy  is obtained from the
smallness of the solar mass splitting, which is $\alpha \Delta$. In 
matter, the mass hierarchy is lifted, but the mass hierarchy suppression
is retrieved in $\sin 2\theta_{12}'$, which is proportional to $\alpha$, and thus,
leads to a mass hierarchy suppression of $J_\mathrm{CP}'$. 

Another interesting point to notice is the factor $1/\CPM$, which leads to an 
MSW-like resonant enhancement of $J_\mathrm{CP}'$ in matter. It can thus be 
expected that the CP-terms $P_{\sin\delta}$ and  $P_{\cos\delta}$ will benefit
from the MSW-resonance in the same way as the leading two neutrino term $P_0$
does.

\section{\boldmath The $\nu_e \rightarrow \nu_\mu$ appearance probability}

Having presented the parameter mapping in matter, it is  now possible to 
start from the ordinary vacuum expressions (eq.~(\ref{3NFORMULA})) in order 
to derive the oscillation probabilities in matter.
The ${J_{ij}^{lm}}'$ as series expansion in $\alpha$ take the following
shape:
\begin{subequations}
\label{JIJLMS}
\begin{eqnarray}
\re {J_{12}^{e\mu}}' &=&  - \cos\delta'  \sin\theta_{12}' \cos^2\theta_{13}' \sin\theta_{13}'  \cos\theta_{23}' \sin\theta_{23}' \nonumber \\ 
                      && - \sin^2\theta_{12}' \cos^2\theta_{23}' + \mathcal{O}(\alpha^3)  \\
\re {J_{13}^{e\mu}}' &=&  - \cos\delta'  \sin\theta_{12}' \cos^2\theta_{13}' \sin\theta_{13}'  \cos\theta_{23}' \sin\theta_{23}'  \nonumber \\ 
                      && - \sin^2 2\theta_{13}' \sin^2\theta_{23}' + \mathcal{O}(\alpha^3) \\
\re {J_{23}^{e\mu}}' &=&  \cos\delta'  \sin\theta_{12}' \cos^2\theta_{13}' \sin\theta_{13}'  \cos\theta_{23}' \sin\theta_{23}' + \mathcal{O}(\alpha^3) \\
\im {J_{12}^{e\mu}}'  =  -\im {J_{13}^{e\mu}}' =  \im {J_{23}^{e\mu}}' &=&  \cos\delta'  \sin\theta_{12}' \cos^2\theta_{13}' \sin\theta_{13}'  \cos\theta_{23}' \sin\theta_{23}'+ \mathcal{O}(\alpha^3)  
\end{eqnarray}
\end{subequations}
Even though in general the calculations were performed only up to order $\alpha$, 
a closer look at
$\alpha^2$-terms proves to be important. Each second term of  
$\re {J_{12}^{e\mu}}'$ in eq.~(\ref{JIJLMS}) is of order $\alpha^2$. Since $\theta_{12}'$
is not suppressed by $\theta_{13}$, these terms give a non-negligible contribution to
the overall oscillation probability. This order $\alpha^2 \sin^0 \theta_{13}$ 
contribution, which will be identified with the $P_3$-term in vacuum (eqs.~(\ref{PROBVACUUM}))
is important for small values of $\theta_{13}$. It is possible to
show without explicit calculation of all order $\alpha^2$-terms of the parameter 
mapping that no further terms of this kind exist. All other $\alpha^2$-terms
in the oscillation probability will at least be suppressed by one power of 
$\theta_{13}$.

Inserting the expression for the mixing parameters in matter together with the
abbreviation $\hat{\Delta} = \Delta \frac{L}{4 E}$ gives the following list of terms
contributing to the oscillation probability $P({\nu_e \rightarrow \nu_\mu})$:
\begin{subequations}
\label{PROBSMATTER}
\begin{eqnarray}
P_0 &=&  \sin^2 \theta_{23} \frac{\sin^2 2\theta_{13}}{\CPM^2} \sin^2(\hat{\Delta} \CPM ) \\
P_{\sin\delta} &=&  \frac{1}{2} \alpha\; \frac{\sin\delta \cos\theta_{13} \sin 2\theta_{12} \sin 2\theta_{13} \sin 2\theta_{23}}{\Ap\CPM\cos\theta_{13}^2} 
 \sin(\CPM \hat{\Delta}) \left[ \cos(\CPM\hat{\Delta}) - \cos((1+\Ap)\hat{\Delta}) \right]  \\
P_{\cos\delta} &=&  \frac{1}{2} \alpha\; \frac{\cos\delta \cos\theta_{13} \sin 2\theta_{12} \sin 2\theta_{13} \sin 2\theta_{23}}{\Ap\CPM\cos\theta_{13}^2}
 \sin(\CPM \hat{\Delta}) \left[ \sin((1+\Ap)\hat{\Delta}) \mp \sin(\CPM\hat{\Delta})  \right]  \\
P_1 &=&  - \alpha\; \frac{1-\Ap \cos 2\theta_{13}}{\CPM^3} \sin^2 \theta_{12} \sin^2 2\theta_{13} \sin^2 \theta_{23}  \hat{\Delta} \sin(2 \hat{\Delta} \CPM) \nonumber \\
     && + \alpha\; \frac{2\Ap(-\Ap + \cos 2\theta_{13})}{\CPM^4} \sin^2 \theta_{12} \sin^2 2\theta_{13} \sin^2 \theta_{23} \sin^2(\hat{\Delta} \CPM)  \\
P_2 &=&  \alpha\; \frac{\mp1 + \CPM \pm \Ap\cos 2\theta_{13}}{2 \CPM^2\Ap \cos^2 \theta_{13}} \cos \theta_{13} \sin 2\theta_{12} \sin 2\theta_{13} \sin 2\theta_{23} \sin^2(\hat{\Delta} \CPM ) \\
P_3 &=&  \alpha^2 \frac{2 \CPM \cos^2 \theta_{23} \sin^2 2\theta_{12}}{\Ap^2 \cos^2\theta_{13}(\mp\Ap + \CPM \pm \cos 2\theta_{13})} \sin^2\left(\frac{1}{2}(1+\Ap\mp\CPM)\hat{\Delta}\right)   
\end{eqnarray}
\end{subequations}
The probability $P({\bar\nu_e \rightarrow \bar\nu_\mu})$ can
be obtained from the probability $P({\nu_e \rightarrow \nu_\mu})$ by flipping 
the sign of the $P_{\sin\delta}$ term.
In all expressions with two possibilities for the sign, 
the upper sign is valid for $\Ap < \cos 2\theta_{13}$ and the lower sign is valid for 
$\Ap > \cos 2\theta_{13}$. The $\Ap$-dependent
pre-factors of $P_1, P_2$, and $P_3$ expanded in $\theta_{13}$ give:
\begin{eqnarray}
\frac{1-\Ap \cos 2\theta_{13}}{\CPM^3} &=&  \pm \frac{1}{(\Ap-1)^2} + \mathcal{O}(\theta_{13}^2) \nonumber \\
\frac{2\Ap(-\Ap + \cos 2\theta_{13})}{\CPM^4} &=&  -\frac{2\Ap}{(\Ap-1)^3} + \mathcal{O}(\theta_{13}^2) \nonumber\\
\frac{\mp1 + \CPM \pm \Ap\cos 2\theta_{13}}{2 \CPM^2\Ap \cos^2 \theta_{13}} &=&   \mathcal{O}(\theta_{13}^2) \nonumber\\
\frac{2 \CPM}{\cos^2\theta_{13}(\mp\Ap + \CPM \pm \cos 2\theta_{13})}  &=&   1 + \mathcal{O}(\theta_{13}^2) \nonumber
\end{eqnarray}
Thus, $P_1$ is quadratic in $\sin \theta_{13}$ and
$P_2$ even of third order in $\theta_{13}$. Therefore, $P_1$ and $P_2$ are negligibly
small compared to $P_{\sin\delta}$ and $P_{\cos\delta}$. The term $P_3$
is important, since it is the only term, which is not suppressed by 
$\theta_{13}$. It was stated before that in some cases the expressions for
the eigenvalues and eigenvectors are not good at the resonance $\Ap = \cos 2\theta_{13}$.
This problem stems from the second order in $\theta_{13}$. On the level of
probabilities, this deficiency is small and only visible in the $P_{\cos\delta}$-term for
large values of $\theta_{13}$. It turns out that neglecting the subleading terms, which
are the source of this problem, gives very accurate results also for $\Ap  = \cos 2\theta_{13}$.
This modification can be applied to both the $P_{\cos\delta}$-term and the $P_{\sin\delta}$-term:
\begin{subequations}
\label{MODIF}
\begin{eqnarray}
P_{\sin\delta} &=&  \alpha\; \frac{\sin\delta \cos\theta_{13} \sin 2\theta_{12} \sin 2\theta_{13} \sin 2\theta_{23}}{\Ap\CPM\cos\theta_{13}^2} 
 \sin \CPM \hat{\Delta} \sin \hat\Delta \sin \Ap\hat\Delta   ~,  \\
P_{\cos\delta} &=&  \alpha\; \frac{\cos\delta \cos\theta_{13} \sin 2\theta_{12} \sin 2\theta_{13} \sin 2\theta_{23}}{\Ap\CPM\cos\theta_{13}^2}
 \sin \CPM \hat{\Delta} \cos \hat\Delta \sin \Ap\hat\Delta  ~.  
\end{eqnarray}
\end{subequations}

\begin{figure}
\begin{center}
\epsfig{file=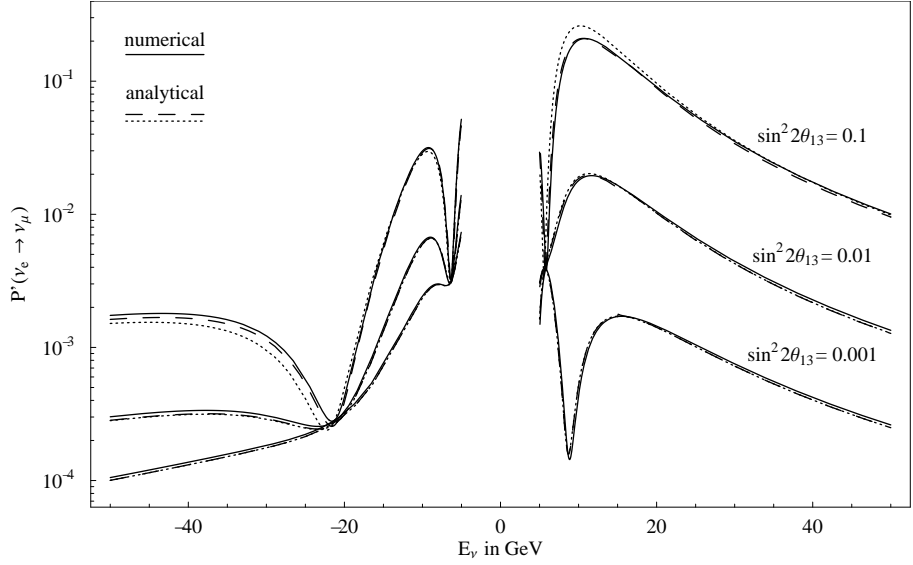,width=12cm}
\end{center}
\caption{Analytical results (dashed and dotted lines) compared to numerical results 
(solid line) for the oscillation probability $P(\nu_e \rightarrow \nu_\mu)$ in
matter ($2.8~\gcc$) as 
function of the neutrino energy. Negative energies correspond to anti-neutrinos.
The dashed line uses the expressions \ref{PROBSMATTER}a,b,c,f. The dotted line
was obtained from equations \ref{PROBSMATTER2}a,b,c,d. The calculation was
performed for the baseline $L$=7000~km with $\delta=0$, bimaximal mixing and three values 
of $\sin^2 2\theta_{13}$ (0.1, 0.01, 0.001). The squared mass differences
are $\dm{31} = 3.2\cdot 10^{-3} \eV^2$ and $\dm{21} = 1 \cdot 10^{-4} \eV^2$.}
\label{FIG:PROBTEST}
\end{figure}

Neglecting all subleading terms in $\theta_{13}$, the relevant terms $P_0$, $P_{\sin\delta}$,
$P_{\cos\delta}$, and $P_3$ take the following simple shapes:
\begin{subequations}
\label{PROBSMATTER2}
\begin{eqnarray}
P_0 &=&  \sin^2 \theta_{23} \frac{\sin^2 2\theta_{13}}{(\Ap-1)^2} \sin^2((\Ap-1)\hat{\Delta})  \\
P_{\sin\delta} &=&   \alpha\; \frac{\sin\delta \cos\theta_{13} \sin 2\theta_{12} \sin 2\theta_{13} \sin 2\theta_{23}}{\Ap(1-\Ap)}
 \sin(\hat{\Delta})\sin(\Ap\hat{\Delta})\sin((1-\Ap)\hat{\Delta})   \\
P_{\cos\delta} &=&   \alpha\; \frac{\cos\delta \cos\theta_{13} \sin 2\theta_{12} \sin 2\theta_{13} \sin 2\theta_{23}}{\Ap(1-\Ap)}
 \cos(\hat{\Delta})\sin(\Ap\hat{\Delta})\sin((1-\Ap)\hat{\Delta})     \\
P_3 &=&  \alpha^2 \frac{\cos^2 \theta_{23} \sin^2 2\theta_{12}}{\Ap^2} \sin^2(\Ap\hat{\Delta})  
\end{eqnarray}
\end{subequations}
It is evident that in the limit of small baselines, $\hat{\Delta} \rightarrow 0$, these expressions converge
to the results in vacuum (eqs.~(\ref{PROBVACUUM}a-d)). A numerical study shows that
the precision loss of eqs.~(\ref{PROBSMATTER2}a-d) compared to eqs.~(\ref{PROBSMATTER}a-f) is 
only relevant for the largest allowed values of $\sin^2 2\theta_{13}$ near the
CHOOZ-bound (0.1). The precision loss is mainly caused by the approximations made in
$P_0$. The term $P_3$ contributes to the overall probability only for small  
$\theta_{13}$, and hence, does not suffer an appreciable accuracy-loss in the form 
given in eq.~(\ref{PROBSMATTER2}d). Figure \ref{FIG:PROBTEST} shows a comparison
of the analytic results obtained here with the results obtained from a numerical
study. Note that the combined contributions from eq.~(\ref{PROBSMATTER}a), eqs.~(\ref{MODIF}a,b) and 
eq.~(\ref{PROBSMATTER2}d) are identical to the result obtained by Cervera~et~al. \cite{GOLDEN} (eq. (16)). 
A similar approach has been discussed in ref.~\cite{FLPR}. 
However, eq.~(16) therein does not cover the case of very small $\theta_{13}$, since
it does not include order $(\dm{21}/\dm{31})^2$ corrections.

\section{Applications}

\subsection{\boldmath Validity region of the low $L/E_\nu$ approximation in matter}

Frequently, the low $L/E_\nu$ limit is used to simplify complex calculations or
derive power laws for neutrino rates. In vacuum, it is well-known
that this approximation is valid for
\begin{equation}
\label{VAL1}
\hat\Delta  \lesssim  1 ~\Rightarrow~ E_\nu \gtrsim 4.0 \GeV 
\left( \frac{\dm{31}}{3.2 \cdot 10^{-3}\eV^2}\right)  
\left( \frac{L}{1000~\mathrm{km}}\right) ~.
\end{equation}
With the use of eqs.~(\ref{PROBSMATTER2}a-d), it is possible to extend this
argument to the presence of matter. Note that in the oscillatory terms,
which are linearized in the small $\hat\Delta$ approximation, there
now also appear the terms $\Ap \hat\Delta$, which must be small. In this 
product, the dependences on the energy $E_\nu$ and the mass squared 
difference $\dm{31}$ cancel. Hence, in addition to relation
(\ref{VAL1}), a direct limit on the baseline $L$, which 
only depends on the matter density $\rho$ is obtained:
\begin{equation}
\label{VAL2}
\Ap \hat\Delta  \lesssim  1 ~\Rightarrow~ L \lesssim 3700~\mathrm{km}  
\left( \frac{\rho}{2.8~\gcc}\right)^{-1} ~.
\end{equation}

\subsection{\boldmath CP-asymmetry in matter at small $L/E_\nu$}

CP-violation studies frequently focus on the fundamental quantity called 
CP-asymmetry $A_\mathrm{CP}$:
\begin{equation}
A_\mathrm{CP} = \frac{P(\nu_e \rightarrow \nu_\mu)-P(\bar\nu_e \rightarrow \bar\nu_\mu)}
{P(\nu_e \rightarrow \nu_\mu)+P(\bar\nu_e \rightarrow \bar\nu_\mu)} ~.
\end{equation}
In vacuum, being proportional to $\sin \delta$, $A_\mathrm{CP}$ is a direct measure 
for intrinsic CP-violation. Since $A_\mathrm{CP}$ is a ratio of probabilities, it
has the important advantage that, on the level of rates, systematic experimental 
uncertainties to a large degree cancel out. However, matter effects also create fake 
CP-asymmetry,
which spoils measurements of the intrinsic CP-violation induced by $\delta$.
The problem to distinguish these two different sources of CP-violation
is often called the ``disentanglement problem''. In a typical long baseline
neutrino experiment, the strength of matter induced CP-effects reaches
the strength of intrinsic CP-effects at baselines around 1000~km. 

Using the above derived approximative solutions for the appearance probability
$P(\nu_e \rightarrow \nu_\mu)$, it is possible to calculate the small $\hat\Delta$
limit of $A_\mathrm{CP}$. For bimaximal mixing ($\theta_{23} = \theta_{12} = \pi/4$)
$A_\mathrm{CP}$ is given by
\begin{equation}
\label{ACP}
A_\mathrm{CP} \approx \frac{2 \hat\Delta \sin 2\theta_{13} \cos\theta_{13}(\alpha \hat\Delta \Ap \cos\delta - 3 \alpha \sin\delta + 2 \hat\Delta \Ap \sin\theta_{13})}
{3 (\alpha^2+ 2\alpha \cos \theta_{13} \sin 2\theta_{13} + \sin^2 2\theta_{13})} \sim \frac{1}{E_\nu} ~.
\end{equation}
The approximation is valid in the regime given by eqs.~(\ref{VAL1}) and (\ref{VAL2}).
This limit is helpful to describe the behavior of 
$A_\mathrm{CP}$ for higher neutrino energies at not too long baselines.
It is interesting to notice that in principle the leading contribution to 
$A_\mathrm{CP}$ in $\hat\Delta$ has its origin in the $\sin\delta$ term. At first
sight, this would suggest to distinguish this intrinsic contribution from 
matter contribution of order  $\hat\Delta^2$ by the energy dependence of 
$A_\mathrm{CP}$. However, taking into account that $\Ap$ itself is proportional
to $E_\nu$, it turns out that all terms in eq.~(\ref{ACP}) have the same energy
dependence $1/E_\nu$. To summarize: In leading order in $\hat\Delta$, 
the CP-asymmetry in matter is proportional to  $1/E_\nu$. 
\begin{figure}[h!]
\begin{center}
\epsfig{file=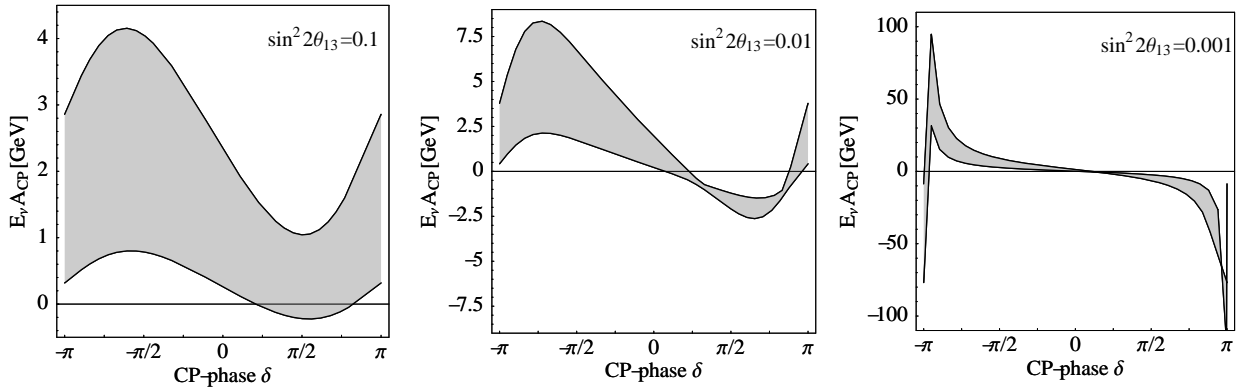,width=\textwidth}
\end{center}
\caption{Dependence of the high energy limit of the CP-asymmetry on the
CP-phase $\delta$ for bimaximal mixing. On the ordinate is plotted the value of 
$E_\nu A_\mathrm{CP}$ in GeV, which should be energy independent
in the low $L/E_\nu$ approximation. The solar mass splitting was
chosen at the upper edge of the LMA-MSW solution $\dm{21} = 1 \cdot 10^{-4} \eV^2$
and the atmospheric mass splitting was varied in the Super-Kamiokande allowed 
90\% confidence interval $3.2\cdot 10^{-3} < \dm{31} < 3.6\cdot 10^{-3} \eV^2$.
The calculation was performed for a baseline of 1000~km.}
\label{FIG:CPASS}
\end{figure}
The coefficient, which describes the $1/E_\nu$-energy dependence of 
$A_\mathrm{CP}$ for high energies is sensitive to both, matter effects
$\Ap$ and intrinsic CP-effects from $\delta$. At high energies, the
quantity $E_\nu A_\mathrm{CP}$ is predicted to be constant in the
energy spectrum and this characteristical quantity
could give direct access to the CP-phase $\delta$. This is 
demonstrated in fig.~\ref{FIG:CPASS}, which shows the value
of $E_\nu A_\mathrm{CP}$ as function of the CP-phase $\delta$ at different 
values of $\sin^2 2\theta_{13}$. Since $E_\nu A_\mathrm{CP}$ does not 
vary with the energy, this simple analysis is to a good approximation 
independent from the energy distribution of the neutrino beam. It is 
of course questionable if, in a real experiment, in the constant regime 
of $E_\nu A_\mathrm{CP}$, there are enough neutrino events to measure. Also 
this method cannot replace a full and detailed statistical analysis
of the complete neutrino energy spectrum.

\subsection{\boldmath Strength of the CP-terms $P_{\sin\delta}$ and $P_{\cos\delta}$}

The two subleading terms (\ref{PROBSMATTER}b) and (\ref{PROBSMATTER}c) currently 
raise considerable interest as they contain information about the CP-phase $\delta$
of the neutrino sector. Today, much effort is spent on the study of CP-violating 
effects in neutrino oscillation experiments \cite{CPPAPERS}. One can try a simple 
approach to this problem by using the here obtained analytic results. It would,
for example, be interesting to know, how strong the information  on $\delta$ 
inherent to the appearance oscillation probability is. To quantify this, one can
look at the relative magnitude of $|P_{\sin\delta} + P_{\cos\delta}|$ 
compared to the statistical fluctuations $\sqrt{P_0 + P_3}$  
in the background signal (provided the errors are Gaussian). 
To obtain statistical meaningful numbers, the estimation should be
performed at the level of event rates expected in a real experiment, e.g. a
neutrino factory long baseline experiment.
Typically, flux times cross sections of a neutrino factory beam \cite{GEER98} 
scales like $E_\nu^3/L^2$. A neutrino factory of $20 \GeV$ muon energy and $10^{20}$
useful muon decays per year produces 54800 $\nu_\mu$-events in a 10~kton detector
at 1000~km distance (assuming measurements in the appearance channel). As a statistical
estimate the following ratio could be chosen:
\begin{equation}
\label{S}
S = \sqrt{54800 \left(\frac{E_\mu}{20\GeV}\right)^3
\left(\frac{L}{1000~\mathrm{km}}\right)^{-2}}
\;\frac{|P_{\sin\delta} + P_{\cos\delta}|}{\sqrt{P_0 + P_3}} 
 ~.
\end{equation}
\begin{figure}[h!]
\begin{center}
\epsfig{file=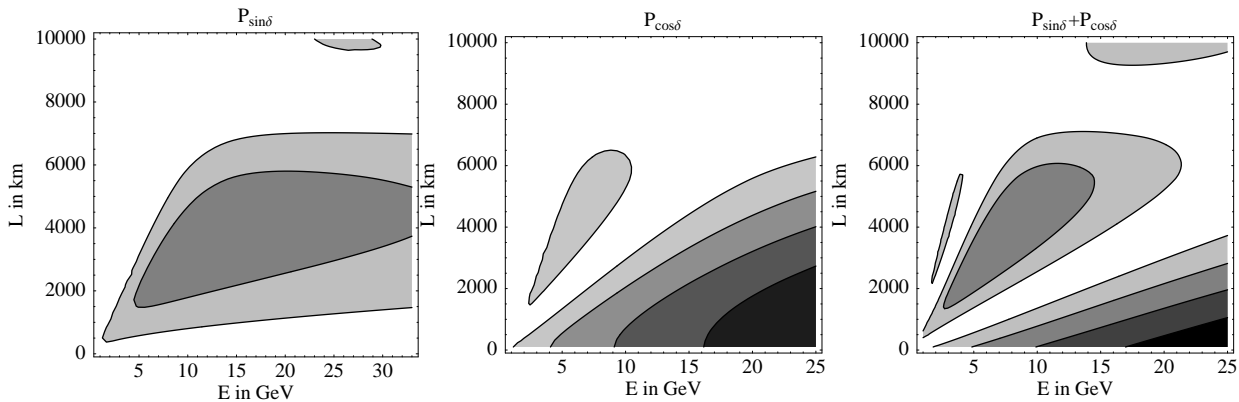,width=\textwidth}
\end{center}
\caption{$1\sigma$, $2\sigma$, $3\sigma$ and $4\sigma$ contour lines
of the quantity $S$ (eq.~(\ref{S})) in the $L-E_\mu$ parameter plane.
Light shading indicates no signal and dark shading indicates strong signal.
The left plot studies only the $P_{\sin\delta}$ term. The plot in the middle
displays the strength of the $P_{\cos\delta}$ term. The right plot, which combines both
terms should give the best approximation to more complex studies. Note
that no energy spectrum was used in this crude model. The calculations were
performed with $\delta = \pi/2$ (left), $\delta = 0$ (middle),  $\delta = \pi/4$ 
(right), bimaximal mixing, and $\sin^2 2\theta_{13} = 0.01$.
The mass squared differences are $\dm{31} = 3.2\cdot 10^{-3} \eV^2$ and $\dm{21} = 1 \cdot 10^{-4} \eV^2$.}
\label{FIG:LE}
\end{figure}

The value of $S$ gives the number of standard deviations (``$\sigma$'s'') at which
the CP-signal is distinct from the ``background''. Figures \ref{FIG:LE} show
the contour lines  $1\sigma$, $2\sigma$, $3\sigma$ and $4\sigma$
of $S$ in the $L-E_\mu$ parameter plane. The plots were produced with a 
running average matter density matched to the baseline $L$. It is interesting
to note that in most of the $L-E_\mu$ parameter space, there is no obvious decrease of the statistical sensitivity to CP-effects 
for increasing beam energy $E_\mu$ as often quoted in the literature. To study this
point in more detail, it is helpful to derive the low $L/E_\nu$ (eq.~\ref{VAL1})
scaling laws for $S$ in the cases $\sin\delta = 1$ and $\cos\delta=1$:
\begin{equation}
S_{\sin\delta} \sim \frac{L}{\sqrt{E_\mu}} \quad \mathrm{and} \quad 
S_{\cos\delta} \sim \sqrt{E_\mu} ~.
\end{equation}
Indeed, for the $P_{\sin\delta}$-term, the statistical sensitivity should 
decrease like $1/\sqrt{E_\mu}$. However, the validity-region of the low $L/E_\nu$
approximation, according to eq.~(\ref{VAL1}), is $E_\nu \gtrsim (4, 12, 20)$~GeV
for $L = (1000, 3000, 5000)$~km. In the left plot of fig.~\ref{FIG:LE} it can be seen 
that roughly at
these energies, $S$ shows a plateau where its maximal value is reached. The
argument in favor of small energies thus only holds for very small baselines
around 1000~km and smaller. The sensitivity to the $P_{\cos\delta}$-term increases like
$\sqrt{E_\mu}$. Hence, in the case of large $\cos\delta$, high beam energies are 
favored to extract information on the CP-phase $\delta$. In conclusion, 
the difference of 
the result presented here and statements being found in the literature has two 
sources. First, usually only the explicitly CP-violating part $P_{\sin\delta}$ of the
oscillation probability is assumed to give the CP-signal\footnote{
Frequently, the need for explicit detection of an asymmetry between the two 
CP-conjugated channels is stressed and matter effects are considered
as background, which prevents such measurements. The attitude taken here is, 
however, different: The goal of any experiment is the limitation of the
allowed parameter space for $\delta$, which does not necessarily presume the
detection of explicit CP-violation. Hence, the $P_{\cos\delta}$ contribution
 has the same status as the  $P_{\sin\delta}$-term and matter effects have
to be included in the theoretical model, which is fitted to the experimental 
data.}. 
Second, the high
energy approximation to the oscillation probabilities  is often applied
without careful consideration of its validity region.

\section{Conclusions}

The purpose of this work was to find approximate analytic expressions
for the neutrino mixing parameters and oscillation probabilities 
in the presence of matter. 
It was stated that being interested in approximate solutions it is difficult
to describe both the solar and the atmospheric resonance at the same time.
Therefore, this work is restricted to energies above the solar resonance according to:
\begin{equation}
|\Ap| \gtrsim |\alpha| ~\Rightarrow~ E_\nu \gtrsim 0.45 \GeV 
\left( \frac{\dm{21}}{10^{-4}\eV^2}\right)  
\left( \frac{2.8~\gcc}{\rho}\right) ~.
\end{equation}
For this regime, the complete parameter mapping (eqs.~(\ref{MAPPING})) 
was given as series 
expansion in the small mass hierarchy parameter $\alpha = \dm{21}/\dm{31}$.
It was shown, that the change of the CP-phase $\delta$ in matter is
triple suppressed by the mass hierarchy, the mixing angle $\theta_{13}$
and by $\theta_{23}$ being close to maximal. Furthermore, it was shown that
in order $\dm{21}/\dm{31}$, the relevant contribution to the parameter mapping 
is the correction of $\theta_{12}$ in matter.
The derived parameter mapping was used to compute the $P(\nu_e \rightarrow \nu_\mu)$ 
appearance oscillation probability in matter. 
Effort was made to find simple solutions, which hold over a wide parameter range 
and are easy to compare with the results known from vacuum oscillation. 
An answer, which in the author's point of view fulfills all these requirements is 
the following  set of terms (eqs.~(\ref{PROBSMATTER2})) contributing to 
$P(\nu_e \rightarrow \nu_\mu)$:
\begin{eqnarray}
P_0 &=&  \sin^2 \theta_{23} \frac{\sin^2 2\theta_{13}}{(\Ap-1)^2} \sin^2((\Ap-1)\hat{\Delta}) \nonumber ~, \\
P_{\sin\delta} &=&   \alpha\; \frac{\sin\delta \cos\theta_{13} \sin 2\theta_{12} \sin 2\theta_{13} \sin 2\theta_{23}}{\Ap(1-\Ap)}
 \sin(\hat{\Delta})\sin(\Ap\hat{\Delta})\sin((1-\Ap)\hat{\Delta})  \nonumber ~, \\
P_{\cos\delta} &=&   \alpha\; \frac{\cos\delta \cos\theta_{13} \sin 2\theta_{12} \sin 2\theta_{13} \sin 2\theta_{23}}{\Ap(1-\Ap)}
 \cos(\hat{\Delta})\sin(\Ap\hat{\Delta})\sin((1-\Ap)\hat{\Delta})  \nonumber ~, \\
P_3 &=&  \alpha^2 \frac{\cos^2 \theta_{23} \sin^2 2\theta_{12}}{\Ap^2} \sin^2(\Ap\hat{\Delta}) \nonumber 
\end{eqnarray}
with $\hat{\Delta} = \dm{31} L / (4 E_\nu)$ and 
$\Ap = A/\dm{31} = 2 V E_\nu/\dm{31}$.
This gives qualitatively good results for baselines at which the oscillation over the 
small (solar) mass squared difference can safely be linearized\footnote{Of course, it is also 
possible to give results, which are not limited by this baseline restriction.
However, this approximation is very helpful to obtain simple results.}:
\begin{equation}
\label{LLIMIT}
\alpha\hat{\Delta} \lesssim 1 ~\Rightarrow~ L \lesssim 8000~\mathrm{km} \left(\frac{E_\nu}{\GeV}\right)  
\left(\frac{10^{-4}\eV^2}{\dm{21}}\right)  ~.
\end{equation}
To obtain high precision results for large values of $\theta_{13}$, it is recommended 
not to neglect subleading $\theta_{13}$ effects. The corresponding terms to 
$P(\nu_e \rightarrow \nu_\mu)$ are given by eqs.~(\ref{PROBSMATTER}a,b,c,f). Results for the 
anti-neutrino channel are always obtained by flipping the signs of $P_{\sin\delta}$ 
and $\Ap$.

Using the derived approximations to the oscillation probability, it was shown 
that from relation (\ref{VAL2}) 
a stringent limit on the baseline $L$ can be derived, up to which the small
$L/E_\nu$ approximation in matter is valid. Then, using this approximation, 
an expression for the CP-asymmetry $A_\mathrm{CP}$ in matter was given, which
demonstrates that, for high neutrino energies, $A_\mathrm{CP}$ is decreasing
proportional to $1/E_\nu$. It was proposed that measuring this energy-dependence
 could help to obtain information on the CP-phase $\delta$. Last, it was demonstrated 
that estimations on the experimental sensitivity to the CP-terms in 
$P(\nu_e \rightarrow \nu_\mu)$ can be given. The here obtained results do not favor low neutrino energies for
the CP-violation search. The reason for the discrepancy between this result and 
statements, which can presently be found in the literature, were discussed. 
These topics were discussed only briefly and mainly 
serve as demonstrations of the applicability of the derived formulas.


\vspace*{7mm}

{\bf Acknowledgments:}

I would like to thank E.~Akhmedov, S.~Bilenky, P.~Huber, M.~Lindner, and T.~Ohlsson for
discussions. This work was supported by the 
``Sonderforschungsbereich~375 f\"ur Astro-Teilchenphysik'' der Deutschen 
Forschungsgemeinschaft.


\newpage

\bibliographystyle{phaip}

\end{document}